\documentclass{article}
\usepackage{amsfonts,amssymb, amsmath}

\usepackage[cp1251]{inputenc}
\usepackage[english]{babel}

\newtheorem{prop}{Proposition}

\def\bq{ \begin{equation}}
\def\eq{ \end{equation}}
\def\ben{ \begin{eqnarray}}
\def\en{ \end{eqnarray}}

\begin{document}


\title{Duffing oscillator and elliptic curve cryptography}
\author{ A.V. Tsiganov\\
andrey.tsiganov@gmail.com\\
Saint-Petersburg University
}%
\date{}
\maketitle
\begin{abstract}
A new approach to discretization of the Duffing equation is presented.  Integrable discrete maps
are obtained by using well-studied encrypting operations in elliptic curve cryptography and, therefore, they
do not depend upon standard small parameter assumption.
 \end{abstract}

\section{Introduction}
\setcounter{equation}{0}
The discretization of dynamical systems in an integrability preserving way has been widely investigated in the last decades. Potentially, it has a great impact in many different areas, such as discrete mathematics, algorithm theory, numerical analysis, statistical mechanics, etc.

The aim of this paper is to discuss new discretizations of the Duffing equation
\bq\label{duff-eq}
\ddot{q}+4Aq+8Bq^3=0\,,
\eq
where $q$ denotes the displacement of the system and $A,B$ are real system constant parameters.
Duffing equations describe many kinds of nonlinear oscillatory systems in physics, mechanics and engineering \cite{mook73}.
Discussion of the known discretizations of the Duffing equations can be found in \cite{map14,map03a,map03b,pots81,pots82,map89,sur89,sur03}.
In order to obtain new integrable discrete maps we will make use standard secret protocols of an elliptic curve cryptography \cite{bos13,hand06}.

In the Hamiltonian approach we start with Hamilton function
\bq\label{duff-ham}
H=p^2+Aq^2+Bq^4\,,\qquad
\eq
which defines Hamiltonian equations
\bq\label{ham-eq}
\dot{q}=\dfrac{\partial H}{\partial p}=2p\,,\qquad \dot{p}=-\dfrac{\partial H}{\partial q}=-2Aq-4Bq^3,
\eq
associated with (\ref{duff-eq}). The corresponding stationary Hamilton-Jacobi equation $H=E$ at $q(t)=x$ and $p(t)=y$ defines
an elliptic curve
\bq\label{duff-ell}
X:\qquad y^2+Ax^2+Bx^4-E=0\,.
\eq
Here $(x,y)$ are abscissa and ordinate  of a point on the projective plane, which we distinguish from  coordinates $(q,p)$ on the phase space. A point $P=(x,y)$ on $X$ corresponds to solution of  Hamiltonian equations  (\ref{ham-eq})  with fixed energy $E$, time $t$ and parameters $A,B$.

In modern elliptic curve cryptography point $(x,y)$ on $X$  plays the role of a message, which can be coded to a cryptogram $(x',y')$, which is another point, using some cryptographic protocol.  It could be  protocol based on a divisor arithmetic \cite{hand06}, post quantum protocol based on isogenies \cite{cost16} and so on.  In any case an encrypting operation
\[\mbox{message}\,(x,y)\to \mbox{cryptogram}\,(x',y')\]
defines a discrete map on the phase space
\[
\mbox{solution}\,(q,p)\to \mbox{solution}\,(q',p').
\]
  We want to study properties of such maps  for the Duffing  equation (\ref{duff-eq}).

 \section{Arithmetic on elliptic curve}
 The field of curve-based cryptography has flourished for the last quarter century after Koblitz  and
Miller independently proposed the use of elliptic curves in public-key cryptosystem in the mid
1980’s. Since then, elliptic curves over finite fields have been used to implement many cryptographic systems and protocols, such
as the Diffie-Hellman key agreement scheme, the elliptic curve variant of the Digital Signature Algorithm,  Bitcoin block chain, etc \cite{bos13,cost12,hand06}.

Our aim is to apply this efficient machinery in the theory of nonlinear dynamical systems and mappings.
One of the main differences is that we use elliptic curves over finite fields    in cryptography and elliptic curves over  phase space of the given dynamical system. The second important difference is related to final aims. In cryptology we have to guarantee security and speed of computing. In the nonlinear dynamical systems theory, we have to study properties of equations that modeled behavior of practical problems that arise in engineering, physics, biology and in many other applications.

We first review the most popular arithmetic  formulae for a generic elliptic curve defined by equation
\[
X:\quad y^2=a_4x^4+a_3x^3+a_2x^2+a_1x+a_0\,.
\]
By adding  two points
\[ (x_1, y_1) + (x_2, y_2) = (x_3, y_3) \]
one gets the third point with the following abscissa and ordinate
\bq\label{add-gen}
x_3=-x_1-x_2-\dfrac{2b_0b_2+b_1^2-a_2}{2b_1b_2-a_3}\,,\qquad\mbox{and}\qquad y_3=-P(x_3)\,,
\eq
where
\[
P(x)=b_2x^2+b_1x+b_0=\sqrt{a_4}(x-x_1)(x-x_2)+\dfrac{(x-x_2)y_1}{x_1-x_2}+\dfrac{(x-x_1)y_2}{x_2-x_1}.
\]
Points $(x_1, y_1)$, $(x_2, y_2)$ and $ (x_3, y_3)$ are the intersection points of elliptic curve $X$ and parabola $y=P(x)$. It allows us to calculate second order polynomial $P(x)$ by using Lagrange interpolation, see classical  \cite{bak97,gr} and modern discussion \cite{cost12,hand06}.

Roughly speaking,  we can consider points $(x_1,y_1)$,  $(x_2,y_2)$ and $(x_3,y_3)$ as message, secret key and cryptogram, respectively.
A more punctual and detailed description of various secret systems based on arithmetic the elliptic curve points can be found in \cite{bos13,hand06,jac09}.

Doubling the point
\[
(x_2,y_2)=[2](x_1,y_1)\,,
\]
is an example of the so-called keyless cryptographic algorithm, which gives rise to cryptogram
\bq\label{doub-gen}
x_2=-2x_1-\dfrac{2b_0\sqrt{a_4}+b_1^2-a_2}{2b_1\sqrt{a_4}-a_3}\,,\qquad\mbox{and}\qquad
y_2=-P(x_2)
\eq
directly from the message $(x_1,y_1)$, i.e. without a secret key.  The points $(x_1, y_1)$ and $ (x_2, y_2)$ are the intersection points of elliptic curve $X$ and parabola $y=P(x)$, where
\[
P(x)=b_2x^2+b_1x+b_0=\sqrt{a_4}(x-x_1)^2+\dfrac{(x-x_1)(4a_4x_1^3+3a_3x_1^2+2a_2x_1+a_1)}{2y_1}+y_1
\]
is the second order polynomial obtained now by Hermite  interpolation \cite{cost12,gr}.

Tripling the point
\[
(x_2,y_2)=[3](x_1,y_1)\,,
\]
where
\bq\label{trip-gen}
x_2= -3x_1-\dfrac{a_3-2b_1b_2}{a_4-b_2^2}\,,\qquad\mbox{and}\qquad
y_2=-P(x_2)\,,
\eq
is also related to the quadratic  polynomial
\[\begin{array}{rcl}
P(x)&=&b_2x^2+b_1x+b_0=-\dfrac{(x-x_1)^2 (4 a_4 x_1^3+3 a_3 x_1^2+2 a_2 x_1+a_1)^2}{8 y_1^3}\\
\\
&+&\dfrac{(x-x_1)\Bigl(x \bigl(6 a_4 x_1^2+3 a_3 x_1+a_2\bigr)-2 a_4 x_1^3+a_2 x_1+a_1\Bigr)}{2 y_1}+y_1\,.
\end{array}
\]
In the similar manner we can consider quadrupling the point and so onc \cite{bos13,hand06}.

Of course, we can extract these formulae either from the original works of Euler, Abel and Jacobi or from the classical textbooks on elliptic functions \cite{bak97,gr}. However, namely cryptology as a computational science has been a driving force behind the arithmetic of
algebraic curves and the other parts of algebraic geometry in the past few decades.  As a result, in cryptology we have formulae, algorithms and even computer programs  prepared for usage \cite{cost12,hand06,har00,jac09}.

\subsection{Integrable maps}
For the Duffing equation (\ref{duff-eq}) elliptic curve $X$  (\ref{duff-ell}) has the extended Jacobi form and we have to put
\[
a_1=a_3=0\,,\qquad a_4=-B\,,\quad a_2=-A\,,\qquad a_0=E
\]
into the standard expressions for addition and multiplications on an integer. Solutions $q(t)$ of the  Duffing equation (\ref{duff-eq}) is  expressed via Jacobi elliptic functions. As an example, for  $B>0$ and $A>-B\alpha^2$ periodic solution of (\ref{duff-eq}) reads as
\[
q(t)=\alpha\,\mbox{cn}\left(2(A+2B\alpha^2)^{1/2}t\,;m\right)\,,\qquad m=\dfrac{\alpha^2B}{A+2\alpha^2B}\,.
\]
For  $B>0$ and $-B\alpha^2<A<-2B\alpha^2$ periodic solution is
\[
q(t)=\alpha\,\mbox{dn}\left(2B^{1/2}t\,;m\right)\,,\qquad m=2\left(1+\dfrac{A}{2\alpha^2}\right)\,.
\]
For $B<0$ and $A>-2B\alpha^2$  periodic solution has the form
\[
q(t)=\alpha\,\mbox{sn}\left(2(A+B\alpha^2)^{1/2}t\,;m\right)\,,\qquad m=-\dfrac{\alpha^2B}{A+\alpha^2B}\,.
\]
Here $\mbox{cn}(z;m)$ and  $\mbox{sn}(z;m)$  are the Jacobi elliptic functions, see \cite{map03a,pots81,pots82}.

We can get integrable discretization of the Duffing equation using these explicit solutions and well-known addition theorems for Jacobi elliptic functions,  for instance
 \[
 \mbox{sn}(X+Y)=\dfrac{\mbox{sn}X\,\mbox{cn}Y\,\mbox{dn}Y+\mbox{sn}Y\,\mbox{cn}X\,\mbox{dn}X}{1-m^2\,\mbox{sn}^2X\,\mbox{sn}^2Y}\,,
 \]
 see discussion in \cite{ts16m}. However, it is more easy and  convenient to apply standard cryptographic algorithms for all the solutions simultaneously.

Let us denote $q_n=q(t_n)$ and $p_n=p(t_n)$, where $t_n$ is time for a $n$-ts step and   $q(t),p(t)$ are solutions of Hamiltonian equations (\ref{ham-eq}). If we substitute coordinates in the phase space   instead of coordinates on the plane
\[x_1=q_n,\qquad y_1=p_n\,,\qquad x_3=q_{n+1}\,,\qquad y_3=p_{n+1}\]
and key on the $n$-th step of coding
\[x_2=\lambda_n\,,\qquad y_2=\mu_n\]
into (\ref{add-gen}) one gets the following discrete map
\bq\label{add-map}
\begin{array}{rlc}
q_{n+1}&=&\frac{q_n-\lambda_n}{B(q_n^2-\lambda_n^2)-\sqrt{-B}(p_n-\mu_n)}
\left(\frac{A+(q_n^2+\lambda_n^2)B}{2}+\frac{\sqrt{-B}(q_n\mu_n-\lambda_n p_n)}{q_n-\lambda_n}+\frac{(p_n-\mu_n)^2}{2(q_n-\lambda_n)^2}\right)\,,\\
\\
p_{n+1}&=&\sqrt{-B}(q_{n+1}-q_n)(q_{n+1}-\lambda_n)+\frac{(q_{n+1}-\lambda_n)p_n}{q_n-\lambda_n}
-\frac{(q_{n+1}-q_n)\mu_n}{\lambda_n-q_n}\,,
\end{array}
\eq
associated with the Duffing equation.

 Similar to \cite{kuz02,ts17d} we can relate secret key $ (\lambda_n,\mu_n)$  with a  discrete time interval $t_{n+1}-t_n$ on $n$-th step of discretization. Because  $ (\lambda_n,\mu_n)$ is an arbitrary point on the elliptic curve $X$ defined by Hamilton-Jacobi equation $H=E$ we can consider abscissa $\lambda_n$ as an  arbitrary number but ordinate is the function of the phase space
\[
\mu_n=\sqrt{H-A\lambda_n^2-B\lambda^4}\,,\qquad H=p_n^2+Aq_n^2+Bq_n^4
\]
and this discrete map is generally two-valued. In elliptic curve cryptography over a finite field  encrypting is always single-valued operation.

\begin{prop}
Mapping (\ref{add-map}) is canonical transformation preserving Hamiltonian (\ref{duff-ham})
\bq\label{nonsymm-ham}
H=p_n^2+Aq_n^2+Bq_n^4
\eq
i.e. it is the area preserving integrable  map.
\end{prop}
The proof is a straightforward calculation.

For the doubling  and tripling  we put
\[x_1=q_n,\qquad y_1=p_n\,,\qquad x_2=q_{n+1}\,,\qquad y_2=p_{n+1}\,.\]
In this case, doubling the point on an elliptic curve  (\ref{doub-gen}) generates  a map
\bq\label{doub-map}
\begin{array}{rcl}
q_{n+1}&=&\dfrac{2Bq_n^4+Aq_n^2+p_n^2}{2\sqrt{-B}q_np_n}\,,\\
\\
p_{n+1}&=&\dfrac{1}{4\sqrt{-B}}\left( \dfrac{4Bq_n^4-p_n^2}{q_n^2}+\dfrac{q_n^2(4B^2q_n^4+4ABq_n^2+A^2)}{p_n^2}\right)\,,
\end{array}
\eq
whereas tripling the point on an elliptic curve  (\ref{trip-gen}) gives rise to another  map
\bq\label{trip-map}
\begin{array}{rcl}
q_{n+1}&=&q_n
-\frac{4 q_n p_n^2 (2 B q_n^4+A q_n^2+p_n^2)}{4 B^2 q_n^8+4 A B q_n^6+8 B q_n^4 p_n^2+A^2 q_n^4+2 A q_n^2 p_n^2+p_n^4}\,,\\
\\
p_{n+1}&=&-p_n+\frac{(q_{n+1}-q_n) \bigl(A(q_{n+1}+q_n)+2B(3q_{n+1}-q_n)q_n^2\bigr)}{2p_n}-\frac{2(q_{n+1}-q_n)^2 (2 B q_n^3+ A q_n)^2}{8 p_n^3}\,.
\end{array}
\eq
\begin{prop}
Mappings (\ref{doub-map})  and (\ref{trip-map})  are  canonical transformations of valence two and three, which  preserve Hamiltonian (\ref{duff-ham})
\[
H=p_n^2+Aq_n^2+Bq_n^4
\]
 i.e. they are integrable  maps doubling and tripling the area on a plane.
\end{prop}
The proof goes via a direct verification.

For Bitcoin cryptographic system $(x,y)$ is the generator point, it is publicly known and is the same for everyone, private key $N$  is the generator multiplier (an integer) and public key $(x',y')$ is the point generated by the private key:
\[
(x',y')=[N](x,y)
\]
For integrable maps $(x,y)=(q,p)$ is some solution of Hamilton-Jacobi equation, $N$ is a valence of canonical transformation preserving Hamilton-Jacobi equation and $(x',y')=(q',p')$ is another solution of the same Hamilton-Jacobi equation.

There are also other secret protocols which  are compositions of cryptographic algorithms and instructions.
In similar manner we can consider various combinations of the discrete maps   (\ref{add-map}), (\ref{doub-map})  and (\ref{trip-map}) for the Duffing equation. Any such combination preserves Hamiltonian  (\ref{duff-ham}) that allows us to compare such discrete maps with known integrable maps also associated with the Duffing equation.

In the numerical integration of nonlinear differential equations, discretization of the nonlinear terms poses extra
ambiguity in reducing the differential equation to a discrete difference equation. For instance, in the framework of the so-called standard-like discretization, the Duffing equation (\ref{duff-eq})
\[
\ddot{q}+4Aq+8Bq^3=0\,,
\]
 can be transformed to the  difference equation
\[
\dfrac{q_{n+1}-2q_n+q_{n-1}}{h^2}+4Aq_n+4B(q_{n+1}+q_{n-1})q_n^2=0\,,
\]
where $h$ is a  discrete time interval.  Then we reduce this equation to the standard expression given by Suris \cite{sur89,sur03} with the mapping function $F(q_n)$
\[
q_{n+1}-2q_n+q_{n-1}=F(q_n)\,,\qquad F(q_n)=\dfrac{-4Aq_n-8Bq_n^3}{h^{-2}+4Bq_n^2}\,.
\]
According   \cite{sur89,sur03} the integrability of this second order difference equation is a sequence of  the integrability of original Duffing equation  (\ref{duff-eq}). In \cite{map03a,map03b} this difference equation is transformed into a standard expression of  two dimensional area preserving map which is integrable, i.e. it admits a nontrivial symmetric invariant integral
\[
\tilde{H}=p_{n+1}^2+Aq_nq_{n+1}+Bq_n^2q_{n+1}^2\,,
\]
which was obtained in the more general case in \cite{sur89,sur03}.

 Discrete maps   (\ref{add-map}), (\ref{doub-map})  and (\ref{trip-map}) preserve nonsymmetric invariant integral (\ref{nonsymm-ham}). It is the main difference between known and new discrete maps associated with the Duffing equation without dumping.

\section{Conclusion}
We obtain new  integrable discrete maps associated with the Duffing oscillator by using addition, doubling and tripling the points of elliptic curves, which are the standard elements of modern elliptic curve cryptography.

 In \cite{ts17d,ts17v,ts17c,ts17p} we apply the same algorithms of  elliptic and hyperelliptic curve cryptography to study discrete versions of the Lagrange top,  H\'{e}non-Heiles system, nonholonomic Veselova and Chaplygin systems, etc. These standard algorithms of hyperelliptic curve cryptography can be also applied to so-called cubic-quintic Duffing oscillator
\[
\ddot{q}+4Aq+8Bq^3+12Cq^5=0\,,
\]
which can be found in the modeling of free vibrations of a restrained uniform beam with intermediate lumped mass, the nonlinear dynamics of slender elastica, the generalized Pochhammer–Chree (PC) equation, the generalized compound KdV equation in nonlinear wave systems, etc.

This system is associated with genus two hyperelliptic curve
\[
X:\qquad y^2-E+Ax^2+Bx^4+Cx^6=0\,.
\]
It is so-called bielliptic curve and, therefore, we have some additional ambiguity to construct explicit and implicit discretizations of the cubic-quintic Duffing oscillator. It can be done using  a well-known computer implementation of the fast arithmetic on genus two hyperelliptic curves  \cite{har00}.  It will be also interesting to apply various  post-quantum  cryptographic algorithms to cubic and cubic-quintic Duffing oscillators.

\end{document}